# Cluster Sliding Ferroelectricity in Trilayer Quasi-Hexagonal C$_{60}$

Xuefei Wang[1#], Yanhan Ren[1#], Shi Qiu[1#], Fan Zhang[1], Xueao Li[1], Junfeng Gao[1]*, Weiwei Gao[1]*, Jijun Zhao[2]
1. Key Laboratory of Material Modification by Laser, Ion and Electron Beams (Dalian University of Technology), Ministry of Education, Dalian 116024, China
2. Guangdong Basic Research Center of Excellence for Structure and Fundamental Interactions of Matter, Guangdong Provincial Key Laboratory of Quantum Engineering and Quantum Materials, School of Physics, South China Normal University, Guangzhou 510006, China
*Email: gaojf@dlut.edu.cn; weiweigao@dlut.edu.cn
#These authors contributed equally to the work.

Electric polarization typically originates from non-centrosymmetric charge distributions in compounds. In elemental crystalline materials, chemical bonds between atoms of the same element favor symmetrically distributed electron charges and centrosymmetric structures, making elemental ferroelectrics rare. Compared to atoms, elemental clusters are intrinsically less symmetric and can have various preferred orientations when they are assembled to form crystals. Consequently, the assembly of clusters with different orientations tends to break the inversion symmetry. By exploiting this concept, we show that sliding ferroelectricity naturally emerges in trilayer quasi-hexagonal phase (qHP) C$_{60}$, a cluster-assembled carbon allotrope recently synthesized. Compared to many metallic or semi-metallic elemental ferroelectrics, trilayer qHP C$_{60}$'s have sizable band gaps and several ferroelectric structures, which are distinguishable by measuring their second-harmonic generation (SHG) responses. Some of these phases show both switchable out-of-plane and in-plane polarizations on the order of 0.2 pC/m. The out-of-plane and in-plane polarizations can be switched independently and enable an easy-to-implement construction of Van der Waals homostructures with ferroelectrically switchable chirality.

Non-centrosymmetric two-dimensional (2D) systems can host a plethora of effects with technological interests, including the Rashba effects [1], bulk photovoltaic effects [2,3], and piezoelectricity [4-6]. Particularly, ferroelectric 2D materials allow for non-volatile switching of electric polarizations in monolayer and few-layer forms. Among them, ferroelectrics with robust out-of-plane (OP) electric polarizations, which are more easily detectable in experiments and implemented in devices, have been pursued by researchers [7-9]. The absence of dangling bonds on the surfaces of 2D ferroelectrics renders them ideal components for building Van der Waals heterostructures with clean and well-defined interface structures, holding promises for applications such as next-generation memory devices [10-14].

Intrinsic 2D ferroelectrics with OP polarizations are rare and mostly compounds. Typically, the chemical bonds between atoms of the same elements favor centrosymmetric structures and charge distributions, which hinder electric polarizations and make elemental 2D ferroelectrics scarce. The discovery of new elemental ferroelectricity can help uncover atypical mechanisms of symmetry breaking. Previous experiments have identified a limited number of elemental ferroelectrics, such as the black phosphorous-like Bi monolayer [15,16]. While black-phosphorus-like bismuth [15] and some of the theoretically proposed 2D elemental materials [16-20] only have in-plane electric polarization, stable elemental ferroelectrics with both OP polarization and a sizable band gap can be more appealing for experimental verification and practical applications [9,13,21,22]. Recently, sliding ferroelectrics offer new opportunities for designing OP ferroelectricity and have been identified in a series of multilayer Van der Waals materials [7,8,23,24], including the semi-metallic multi-layer graphene (with four or more layers) [25-27]. The semi-metallic nature of graphene makes it challenging to apply the standard piezoresponse force microscopy to experimentally confirm the ferroelectricity [28].

In order to expand the family of low-dimension ferroelectrics, we propose to examine cluster-assembled materials. Compared to atomic crystals, systems that consist of interlinked clusters are more likely to break the inversion symmetry. For example, despite $C_{60}$ being known as one of the most stable and symmetric molecules, it is still less symmetric than real atoms, which have perfect spherical symmetry. In few-layer forms, two-dimensional materials consisting of $C_{60}$ with different orientations are more likely to break the inversion symmetry compared to atomic crystals [29,30]. A recently synthesized 2D polymeric fullerene network is one of the ideal candidates for constructing cluster sliding 2D ferroelectrics [31-33]. By exploiting this observation, we demonstrate that trilayer quasi-hexagonal phase $C_{60}$'s have a variety of polar stacking configurations, which are stable elemental ferroelectrics with both sizable band gaps and out-of-plane electric polarizations. These phases can be conveniently distinguished through second-harmonic-generation (SHG) measurements[34,35]. Some trilayer qHP $C_{60}$ have decoupled in-plane and out-of-plane polarizations, which can lead to an easy-to-implement scheme for building chiral vdW homostructures with ferroelectrically switchable chirality.

**Results**

**Stacking modes of bilayer and trilayer qHP $C_{60}$**

In experiments, 2D quasi-hexagonal phase (qHP) and quasi-tetragonal phase (qTP) $C_{60}$ have both been successfully synthesized [31-33]. We choose to investigate the qHP $C_{60}$ since it demonstrates better stability [30,36] than the qTP $C_{60}$ and can be fabricated in both monolayer and few-layer forms. In qHP $C_{60}$, fullerenes adopt two possible orientations[30-32], which can be distinguished from the highlighted pentagons on the top of a $C_{60}$, as shown in Fig. 1 (a).

The monolayer qHP $C_{60}$ has a centrosymmetric structure. In stable few-layer forms, a fullerene tends to align with the center of the triangle consisting of three nearest-neighboring fullerenes from the adjacent layer, as observed in experiments [31,32]. There are two distinct bilayer configurations (labeled as AB and AB') of qHP $C_{60}$, which are both centrosymmetric, as shown schematically in Fig. 1 (b) and (c). In both cases, the bottom layer is shifted relative to the top layer by approximately $b/3$, where $b$ is the lattice constant along the $b$-direction.

Next, to examine whether trialyer qHP $C_{60}$ can be noncentrosymmetric, we enumerate all possible trilayer qHP $C_{60}$'s and identify six different stacking structures. In our naming convention, we use letters A, B, C, A', B', and C' to label the relative positions of different layers. The top configuration layer is fixed to be A in order to avoid counting duplications. A "B" layer stands for it is shifted by approximately -b/3 with respect to the "A" layer. Similarly, a "C" layer stands for the layer shifts by -2b/3 with respect to the "A" layer. For an "A'" layer, each $C_{60}$ cluster vertically aligns with and has the opposite orientation with respect to a corresponding $C_{60}$ of the "A" layer. The same convention applies to "B'" and "C'" layers. The schematic plots of all six trilayer stacking modes are presented in Supplemental Fig. S1 and S2. Among them, two stacking configurations (AB'A', AB'C') possess both out-of-plane and in-plane polarizations; two stacking configurations have only in-plane polarizations; and the other two are centrosymmetric. The stacking configurations, point group, and electric-polarization directions are summarized in Table 1. In the following discussion, we focus on the trilayer phases with out-of-plane polarizations.

Fig. 1 (d) schematically depicts the non-centrosymmetric AB'A' stacking configuration with both IP and OP polarization. To better illustrate the origin of its polarization, we calculated the charge density differences $\Delta\rho(\mathbf{r})$ between qHP $C_{60}$ layers in the AB'A' stacking configuration and free-standing qHP $C_{60}$ monolayers. As shown in Fig. 1 (e), due to the weak interlayer interaction, the redistributed charge density mostly occurs at the boundary of the Van der Waals gap between adjacent layers and evidently reveals the asymmetric shape in the AB'A' structure. In addition, the planar-averaged distribution $\Delta\rho(z) = \int \Delta\rho(\mathbf{r}) dxdy$ along the z-axis quantitatively demonstrates symmetry broken, giving rise to the out-of-plane polarization (see Supplemental Fig. S3 for more details).

**Sliding ferroelectricity of trilayer qHP $C_{60}$**

As shown in Fig. 1 (f), the energy differences between different stacking configurations are less than 0.1 meV/atom, which is beyond the chemical accuracy of DFT calculations. Accordingly, one would expect these stacking configurations to be nearly degenerate ground-state structures and can simultaneously occur in practical situations. With the Berry phase formulation of macroscopic polarization, the calculated out-of-plane polarizations of AB'A' and AB'C' phases are 0.22 and 0.25 pC/m, respectively. This is on the same order of magnitude as the out-of-plane polarization of tetralayer graphene [25], InSe [8], and many others [21]. The ABA and AB'A structures exhibit in-plane polarizations of 0.17 pC/m and 0.18 pC/m, respectively, but no out-of-plane polarization. DFT calculations reveal trilayer qHP $C_{60}$ with different stacking configurations have a Kohn-Sham band gap of around 0.84 eV (as shown in Supplemental Fig S5). Compared to metal or semi-metallic ferroelectrics [25,28,37], the finite band gap and OP polarizations of trilayer qHP $C_{60}$ make it easier to measure the electric polarization in experiments and more suitable for diverse applications.

As different tri-layer stacking configurations have similar energies and can both appear in experiments, it is important to find an effective means to distinguish them. SHG response is a well-established technique for identifying layer stacking of 2D materials with variations of centrosymmetric structures [34,38]. Fig. 2 (a) shows the calculated frequency-dependent second-order nonlinear susceptibility tensor component $\chi^{(2)}_{yyy}$ of AB'A' and AB'C' structures. The highest peak of $\chi^{(2)}_{yyy}$ for the AB'C' structure corresponds to a photon energy of 1.213 eV (equivalent to a wavelength of 1022 nm), while the highest peak for the AB'A' structure is located at 1.329 eV (corresponding to 933 nm). The dependencies of other second-order nonlinear susceptibility tensor components are presented in Supplemental Material Fig. S6. Under an incident light with a wavelength of 1022 nm and an incident angle of $\theta = 0°$, the SHG responses of two trilayer structures are plotted as a function of polarization angle $\phi$ in Fig. 2 (b) and (c) [39,40]. The maximum parallel components $I_{\parallel}(\phi)$ of both AB'A' and AB'C' configurations appear at $\phi = 90°$, but with a substantial difference in their magnitude. Additionally, the vertical component $I_{\perp}(\phi)$ of AB'A' peaks at $\phi = 0°$, with its maximum value comparable to the parallel component. In contrast, the vertical component intensity of AB'C' is only one-twentieth of its parallel component. Clearly, SHG responses can serve as a useful tool for distinguishing trilayer qHP $C_{60}$ stacking configurations.

Similar to other sliding ferroelectrics [31-33], tri-layer $C_{60}$ can be transformed between different stacking configurations by relatively translational displacements of $C_{60}$ layers, i.e., interlayer sliding. For example, a translational displacement of the middle layer of AB'A' stacking structure by $(\boldsymbol{a} + \boldsymbol{b})/2$ directly transforms it into ABA' structure. One should note that ABA' is physically equivalent to the A'B'A structure and has an out-of-plane polarization opposite to that of AB'A'. Consequently, the polarization reversing process is achieved under the process depicted in Fig. 3 (b). The calculated energy barrier required to achieve this transition is about 7 meV/atom, which is on the same order of magnitude as the sliding energy barrier of tetra-layer graphene (about 4 meV/unit cell) [25]. Moreover, the symmetry properties of the AB'A' structure indicate that a displacement of the middle layer by $(-\boldsymbol{a} + \boldsymbol{b})/2$, $(\boldsymbol{a} - \boldsymbol{b})/2$, or $-(\boldsymbol{a} + \boldsymbol{b})/2$ can also transform it to the same structure. This can be illustrated by the energy landscape as a function of different translational movements of the middle layer (Supplemental Fig. S7).

Notably, the energy barrier can be reduced when the polarization reversing process goes through meta-stable stacking configurations. For example, Fig. 3 (b) schematically illustrates the transition process AB'A'→ABA→ABA' (=A'B'A). One may note that the intermediate state ABA is non-polar and equivalent to A'B'A'. This process may be understood by first sliding the top layer $C_{60}$ by $(\boldsymbol{a} + \boldsymbol{b})/2$, followed by sliding the bottom layer $C_{60}$ by $(a + b)/2$. Similarly, another low-energy transition path AB'A'→AB'A→ABA' (=A'B'A) is accomplished by first shifting the bottom layer and then the top layer. The energy barriers for these two transition paths are around 3 meV/atom.

While the AB'A' structure resembles the Bernal stacking order of graphene, the AB'C' structure is more similar to the rhombohedral stacking configuration of graphene [26]. Consequently, the polarization reversal process of AB'C' is distinct from that of AB'A' and can be achieved by sliding the bottom layer and middle layer by $a/2 + b/3$ and $b/3$, as shown in Fig. 3 (d). This transition path leads to the final structure ACB', which is equivalent to C'B'A. Even though this transition path involves a simultaneous movement of two layers, the energy barrier is only around 2 meV/atom, which is lower than many reported 2D ferroelectrics [8,41-43]. Similar to the AB'A' structure, the polarization reversal path of AB'C' can be achieved by passing through a meta-stable intermediate state, such as ACA. This can be accomplished by shifting the top layer by $-a/2 - b/2$, followed by sliding the bottom layer by $a/2 + b/2$ to get ACB' (= C'B'A, which has a lower energy barrier of ~1 meV/atom.

**Transition between different ferroelectric states and switchable chirality**

In addition to the reversal of polarization, switching between different polarization states can also be effectively accomplished by interlayer sliding [44]. We calculated the transition energy barriers between four stacking configurations with various out-of-plane and in-plane polarizations, as shown in Fig. 4. Overall, the energy barriers required for transitioning between different polarization states are close to those required to reverse the direction of polarization. For instance, the energy barriers for transitions AB'C'→AB'A', AB'A'→AC'A', and AC'A'→AC'B' are 0.8, 1.7, and 0.8 meV/atom, respectively.

Notably, the AB'A' and AC'A' configurations are related by a rotation around the c-axis by 180° and thus physically equivalent. Similarly, AB'C' and AC'B' structures are physically equivalent. As a result, AB'A' and AC'A' structures (or AB'C' and AC'B' structures) have the same OP polarization but opposite IP polarizations. In other words, the out-of-plane and in-plane polarizations can be reversed independently in trilayer qHP $C_{60}$ systems. This feature is also found in the ferroelectric multi-layer graphene [cite], but distinct from monolayer $In_2Se_3$ [9] and many others [21], whose OP polarization direction is coupled with the IP polarization direction.

The decoupling of IP and OP polarization in tri-layer qHP $C_{60}$ allows one to construct Van der Waals structures with ferroelectrically switchable chirality. It is noteworthy that intrinsic chirality in 2D Van der Waals materials is uncommon and has been pursued [45]. Chirality in low-dimension materials can be artificially introduced by twisting [46,47], kirigami [48,49], adsorption of chiral molecules [50,51], and other methods [52,53], which typically do not allow for nonvolatile switching of chirality through electric fields. To demonstrate the concept of switchable chirality, we consider two slabs of AB'A' trilayers that are rotated by $\theta = 90°$ with each other, as shown in Fig. 5. In practice, the rotation angle $\theta$ can also be any finite value ranging between 0 to 180°. As shown in Fig. 5, the top A'B'A' trilayer has an OP polarization $\mathbf{P}_{OP1}$ pointing along the $z$-direction and an IP polarization $\mathbf{P}_{IP1}$ pointing along the $x$-direction, while the bottom A'B'A' trilayer has an OP polarization $\mathbf{P}_{OP2}$ pointing along the $z$-direction and an IP polarization $\mathbf{P}_{IP2}$ pointing along the $y$-direction. Such hexa-layer configuration has a total polarization given by $(\mathbf{P}_{IP1}, \mathbf{P}_{IP2}, \mathbf{P}_{OP1} + \mathbf{P}_{OP2})$, that forms a basis with right-handed chirality. The chirality can be readily switched by applying an out-of-plane electric field that reverses the OP polarization while keeping the IP polarization unchanged. This leads to a new configuration with polarization $(\mathbf{P}_{IP1}, \mathbf{P}_{IP2}, -\mathbf{P}_{OP1} - \mathbf{P}_{OP2})$, which has left-handed chirality. As we noted, the key for this ferroelectric chirality switching is to decouple the OP and IP polarizations, i.e., switching the OP polarization independently.

**Discussion**

In conclusion, based on symmetry analysis and first-principles calculations, a few non-centrosymmetric stacking modes of trilayer qHP $C_{60}$ with out-of-plane or in-plane electrical polarizations are predicted. The energy barriers for switching the electrical polarizations are comparable with those of other sliding ferroelectrics. Such unconventional elemental ferroelectricity originates from the orientational dependent structure of $C_{60}$ clusters. The multiple ferroelectric stacking states of trilayer qHP $C_{60}$ can be clearly detected and distinguished via second harmonic generation measurement. We also propose an easy-to-implement scheme that takes advantage of the decoupled in-plane and out-of-plane polarizations to construct Van der Waals few-layer systems with non-volatile switchable chirality. Moreover, sliding ferroelectricity is not limited to trilayer systems. It can be generalized to few-layer qHP $C_{60}$ with more than three layers. For instance, we enumerate tetralayer qHP $C_{60}$ configurations (as shown in Supplemental Table S1). Among them, 11 tetralayer stacking configurations are centrosymmetric. In the future, it would be interesting to investigate the sliding ferroelectricity of tetra- or penta-layer qHP $C_{60}$ or other cluster-assembled systems. Furthermore, the interstitial space between $C_{60}$ cages implies the possibility of intercalating transition metal ions, which may in turn, introduce multiferroicity to the system.

## Methods

### Computational details

The calculations were based on pseudopotential density functional theory (DFT) implemented in the Quantum Espresso (QE) package [54,55]. Projector Augmented Wave (PAW) pseudopotentials which are taken from the SSSP library[56] and the Perdew–Burke–Ernzerhof (PBE) [57] exchange-correlation functional with DFT-D3 dispersion correction [58] were used. A vacuum layer of 30 Å was adopted to isolate the artificial interaction between periodic images in the $z$-direction. For structural relaxation, the convergence thresholds of force and energy were $1 \times 10^{-3}$ Ryd/Bohr and $1 \times 10^{-5}$ Ryd, respectively. The electron wave function was represented using the plane-wave basis set, with a kinetic-energy cutoff of 50 Ry. For the Brillouin zone (BZ) sampling, a 4×2×1 Monkhorst-Pack k-point grid was employed [59]. Electronic polarization was calculated using the Berry phase formulation of macroscopic polarization [60,61]. The energy barriers of ferroelectric switching pathways were obtained using the nudged elastic band (NEB) method [62]. The Second-harmonic generation (SHG) susceptibility was calculated with the independent particle approximation, using a self-developed code interfaced with the QE package[39,40,63].

### Data Availability

The authors declare that the main data supporting the findings of this study are available within the article and its Supplementary Information files.

### Code Availability

### Acknowledgement

We acknowledge the support by the National Natural Science Foundation of China (12104080, 12474221, 12374253, and 12074053). Computational resources are provided by the National Supercomputer Center at Wuzhen.

**Author contributions:** WG conceived and designed the study. XW performed density-functional theory calculations, analyzed the results, plotted the figures, and wrote the first draft of the manuscript. SQ developed the code for second-harmonic generation (SHG) responses calculations. YR and SQ performed the SHG responses calculations and data analysis. WG, JG, and JZ supervised the research and finished the final version of the paper. All authors participated in discussing and editing the manuscripts.

**Competing interests:** the Authors declare no Competing Financial or Non-Financial Interests.

| | |
|---|---|
| | doi:10.1103/PhysRevLett.131.186201 (2023). |
| 30 | Li, X., Zhang, F., Wang, X., Gao, W. & Zhao, J. Rich structural polymorphism of monolayer polymeric $C_{60}$ from cluster rotation. *Phys. Rev. Mater.* **7**, 114001, doi:10.1103/PhysRevMaterials.7.114001 (2023). |
| 31 | Hou, L. *et al.* Synthesis of a monolayer fullerene network. *Nature* **606**, 507-510, doi:10.1038/s41586-022-04771-5 (2022). |
| 32 | Meirzadeh, E. *et al.* A few-layer covalent network of fullerenes. *Nature* **613**, 71-76, doi:10.1038/s41586-022-05401-w (2023). |
| 33 | Wang, T. *et al.* Few-Layer Fullerene Network for Photocatalytic Pure Water Splitting into $H_2$ and $H_2O_2$. *Angew. Chem. Int. Ed.* **62**, e202311352, doi:https://doi.org/10.1002/anie.202311352 (2023). |
| 34 | Hsu, W.-T. *et al.* Second Harmonic Generation from Artificially Stacked Transition Metal Dichalcogenide Twisted Bilayers. *ACS Nano* **8**, 2951-2958, doi:10.1021/nn500228r (2014). |
| 35 | Liu, Y. *et al.* Stacking Faults Enabled Second Harmonic Generation in Centrosymmetric van der Waals $RhI_3$. *ACS Nano* **18**, 17053-17064, doi:10.1021/acsnano.4c03562 (2024). |
| 36 | Peng, B. Stability and Strength of Monolayer Polymeric $C_{60}$. *Nano Lett.* **23**, 652-658, doi:10.1021/acs.nanolett.2c04497 (2023). |
| 37 | Fei, Z. *et al.* Ferroelectric switching of a two-dimensional metal. *Nature* **560**, 336-339, doi:10.1038/s41586-018-0336-3 (2018). |
| 38 | Shan, Y. *et al.* Stacking symmetry governed second harmonic generation in graphene trilayers. *Sci. Adv.* **4**, eaat0074, doi:10.1126/sciadv.aat0074 (2018). |
| 39 | Li, S.-Q. *et al.* Dramatically Enhanced Second Harmonic Generation in Janus Group-III Chalcogenide Monolayers. *Adv. Opt. Mater.* **10**, 2200076, doi:https://doi.org/10.1002/adom.202200076 (2022). |
| 40 | Zhang, A. *et al.* Robust Type-II Band Alignment and Stacking-Controlling Second Harmonic Generation in GaN/ZnO vdW Heterostructure. *Laser Photonics Rev.* **18**, 2300742, doi:https://doi.org/10.1002/lpor.202300742 (2024). |
| 41 | Ding, N. *et al.* Phase competition and negative piezoelectricity in interlayer-sliding ferroelectric $ZrI_2$. *Phys. Rev. Mater.* **5**, 084405, doi:10.1103/PhysRevMaterials.5.084405 (2021). |
| 42 | Wang, Z., Gui, Z. & Huang, L. Sliding ferroelectricity in bilayer honeycomb structures: A first-principles study. *Phys. Rev. B* **107**, 035426, doi:10.1103/PhysRevB.107.035426 (2023). |
| 43 | Ma, X., Liu, C., Ren, W. & Nikolaev, S. A. Tunable vertical ferroelectricity and domain walls by interlayer sliding in β-$ZrI_2$. *npj Comput. Mater.* **7**, 177, doi:10.1038/s41524-021-00648-9 (2021). |
| 44 | Meng, P. *et al.* Sliding induced multiple polarization states in two-dimensional ferroelectrics. *Nat. Commun.* **13**, 7696, doi:10.1038/s41467-022-35339-6 (2022). |
| 45 | Zhu, H. & Yakobson, B. I. Creating chirality in the nearly two dimensions. *Nat. Mater.* **23**, 316-322, doi:10.1038/s41563-024-01814-2 (2024). |
| 46 | Li, G. *et al.* Observation of Van Hove singularities in twisted graphene layers. *Nat. Phys.* **6**, 109-113, doi:10.1038/nphys1463 (2010). |
| 47 | Carr, S. *et al.* Twistronics: Manipulating the electronic properties of two-dimensional layered structures through their twist angle. *Phys. Rev. B* **95**, 075420, doi:10.1103/PhysRevB.95.075420 (2017). |
| 48 | Liu, Z. *et al.* Nano-kirigami with giant optical chirality. *Sci. Adv.* **4**, eaat4436, |

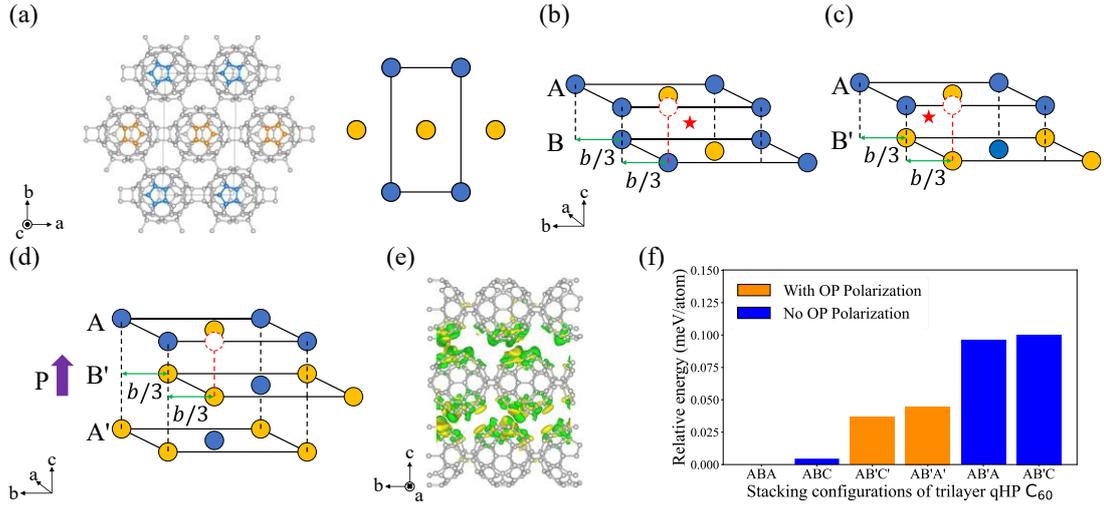

FIG. 1. (a) (left) The structural model of monolayer qHP $C_{60}$. The blue and yellow pentagons are used to highlight $C_{60}$ clusters in two different orientations. (right) A simplified illustration of the qHP $C_{60}$ monolayer, where the blue and yellow spheres represent two different orientations of the $C_{60}$. Bilayer qHP $C_{60}$ in the (b) AB and (c) AB' stacking configuration, with the inversion center labeled by the red star. (d) Trilayer $C_{60}$ with AB'A' stacking configuration; (e) The isosurface plot (with isosurface value of $\pm 6.5 \times 10^{-5}$ $e$/Bohr$^3$) of the differential charge density of the AB'A' structure, where yellow and green isosurfaces indicate electron accumulation and depletion due to non-centrosymmetric layer stacking; (f) The relative energies of 6 stacking configurations and the out-of-plane polarization of the polar structure.

Table 1. The symmetry properties and polarization directions of 6 different trilayer stacking configurations.

| Stacking configurations | Point Group | Out-of-plane (OP) polarization | In-plane (IP) polarization |
|---|---|---|---|
| AB'A', AB'C' | m | ✓ | ✓ |
| ABA, AB'A | mm2 | × | ✓ |
| ABC, AB'C | 2/m | × | × |

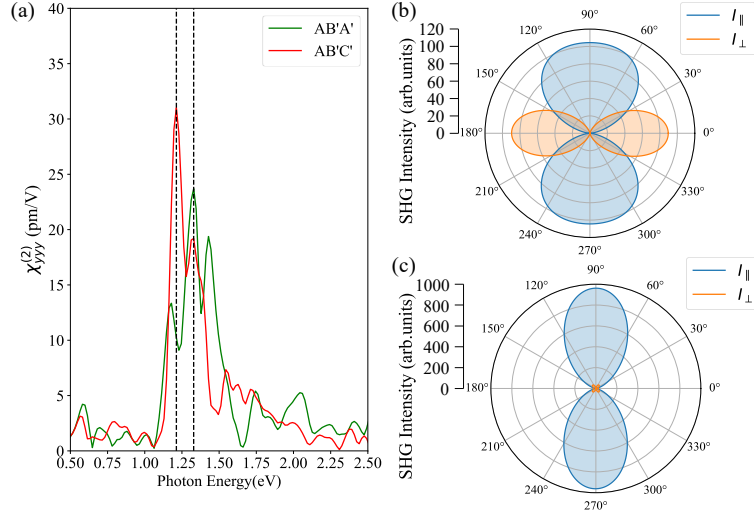

FIG.2 (a) The frequency-dependent SHG susceptibility tensor components $\chi^{(2)}_{yyy}$ of AB'A' and AB'C'; (b) the polarization angle-dependent SHG response intensity of AB'A' stacking configuration; (c) the polarization angle-dependent SHG response of AB'C' stacking configuration. $I_\parallel$ and $I_\perp$ are the parallel and perpendicular components of the SHG response intensity, respectively.

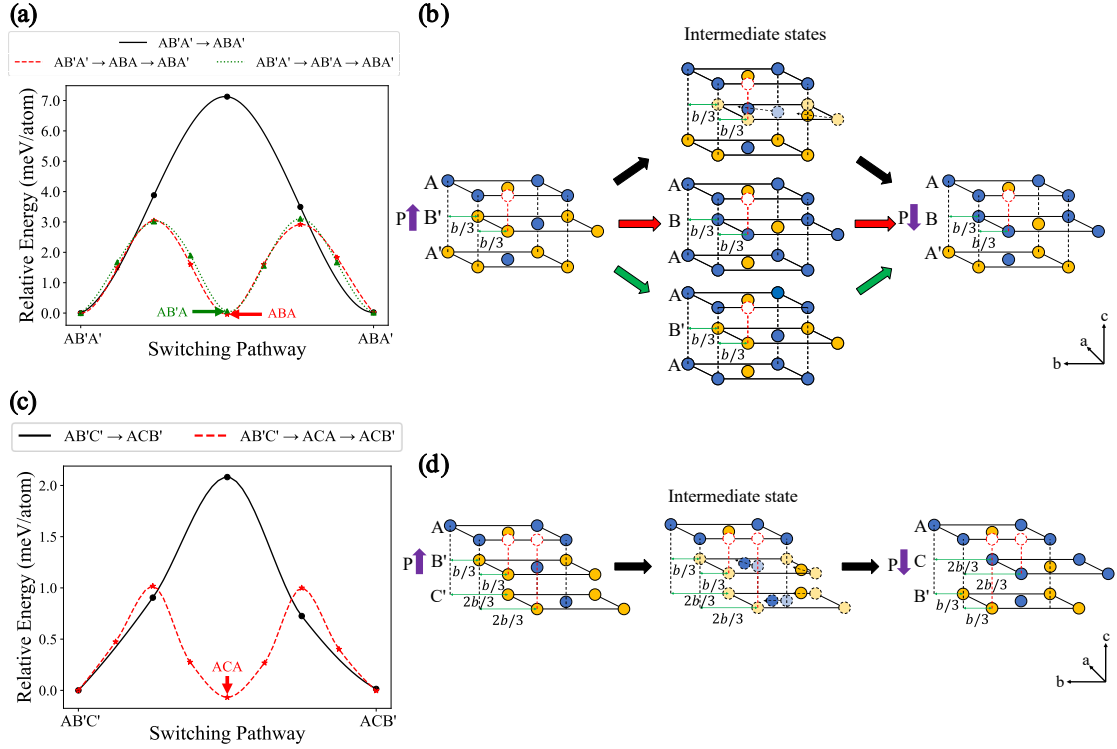

FIG. 3. (a) Energy barriers and (b) schematic diagrams if different polarization-switching paths of the AB'A' configuration; (c) Energy barriers and (d) schematic diagrams of different polarization-switching path of the AB'C' configuration.

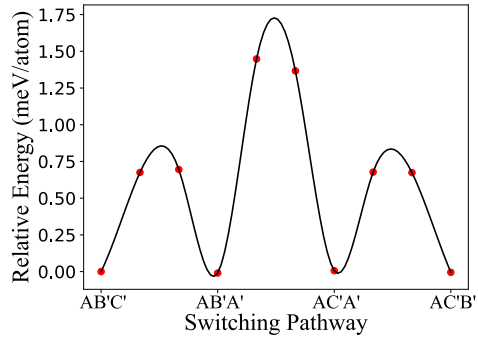

FIG. 4. The energy barriers for the transition between four stacking configurations that have both OP and IP polarizations.

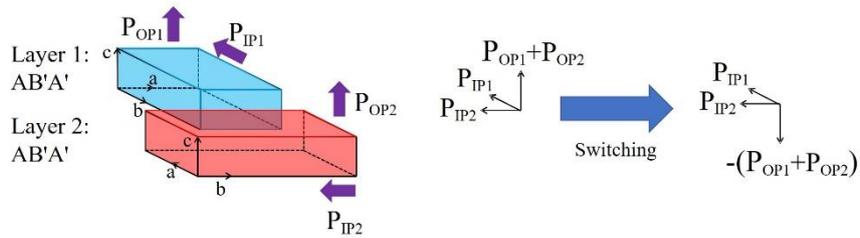

FIG. 5. (left) A schematic plot for a hexa-layer qHP $C_{60}$ system with ferroelectrically switchable chirality. (right) The electric polarization before and after applying a vertical electric field to switch the OP polarization and chirality of the system.